# Hot Phonons in an Electrically Biased Graphene Constriction

*Dong-Hun Chae*[*][†]*, Benjamin Krauss[†], Klaus von Klitzing, and Jurgen H. Smet**

Max-Planck-Institute for Solid State Research, Heisenbergstrasse 1, 70569 Stuttgart, Germany

[*] To whom correspondence should be addressed. E-mail: D.Chae@fkf.mpg.de or J.Smet@fkf.mpg.de

[†] D.Chae and B.Krauss contributed equally to this work.

Phonon carrier interactions can have significant impact on device performance. They can be probed by measuring the phonon lifetime, which reflects the interaction strength of a phonon with other quasi-particles in particular charge carriers as well as its companion phonons. The carrier phonon and phonon-phonon contributions to the phonon lifetime can be disentangled from temperature dependent studies. Here, we address the importance of phonon carrier interactions in Joule-heated graphene constrictions in order to contribute to the understanding of energy dissipation in graphene based electronic devices. We demonstrate that gapless graphene grants electron phonon interactions uncommon significance in particular at low carrier density. In conventional semiconductors, the bandgap usually prevents the decay of phonons through electron-hole generation and also in metals or other semimetals the Fermi temperature is excessively large to enter the regime where electron phonon coupling plays such a dominant role as in graphene in the investigated phonon temperature regime from 300 to 1600 K.

The prospect of carbon based electronics has triggered a strong interest in understanding the scattering processes of charge carriers supporting current flow through these devices. High currents may induce a significant overpopulation of phonons[1-6]. These phonon modes may in turn impede transport and enhance scattering with electrons as a result of strong electron-phonon interactions[1-3, 6-8]. For instance, Yao et al.[1] have reported a dramatic drop in the ballistic conductance in carbon nanotubes under large applied bias voltages which would deteriorate the performance of carbon nanotube based interconnects. Here we exploit the local generation of hot phonons in Joule-heated graphene constrictions with high energy electrons to investigate phonon decay mechanisms and especially the electron phonon scattering as a function of temperature. The main tool is scanning confocal Raman spectroscopy of the in-plane carbon stretching mode (G-mode). Linewidths associated with phonons immediately reveal scattering rates and the Stokes and anti-Stokes counterparts allow extracting the phonon population number and its spatial dependence along the constriction. The temperature dependence of the phonon linewidth is non-monotonic and decreases with temperature for some temperature range at sufficiently low carrier densities. This is very distinct from the monotonically increasing linewidth of optical phonons due to anharmonic phonon-phonon scattering, which dominate in the large majority of other materials[9-11] in the investigated temperature regime.

A scanning electron microscope picture of a typical device geometry is shown in the left inset of Figure 1a. It consists of a monolayer graphene flake with an etched constriction at its center with a width of 0.6 μm and a length of 1.5 μm. This geometry offers the advantage of concentrating the heat dissipation locally at the constriction. Graphene flakes were peeled off from highly-oriented pyrolytic graphite with adhesive tape and transferred on a degenerately-doped Si substrate with a 300 nm thermal $SiO_2$ on top. Monolayer graphene was identified using the contrast difference on 300 nm $SiO_2$ with optical microscopy[12]. The line shape and position of the 2D peak from Raman spectroscopy was utilized as an unequivocal check[13]. The 2D Raman spectrum of the measured graphene is depicted in the Supporting Information. The constrictions were produced by first patterning an etch mask with electron-beam lithography and evaporation of 25 nm of Al. The uncovered graphene was etched away with the help of an Ar plasma. The Al etch mask was subsequently dissolved in a 1 Mol KOH solution. Finally, Cr/Au (5nm/50 nm) electrodes were deposited after a last e-beam lithography step.

The left inset of Figure 1a depicts schematically the scanning confocal Raman setup. Raman measurements were performed with a bias voltage applied across the graphene constriction in a backscattering geometry using a He-Ne laser at 632.8 nm with a diffraction-limited spot size of approximately 500 nm. The incident laser beam is deflected by a piezo-driven mirror to obtain a scanning confocal optical image of the sample in order to find the constriction and precisely position the laser spot. With the same method, spatially-resolved Raman spectra were acquired. All measurements were carried out at room temperature in vacuum with a pressure below $3 \times 10^{-6}$ mbar to avoid unintentional contamination. Stokes and anti-Stokes Raman spectra were measured simultaneously using a notch filter. The power of the incident laser light was optimized and fixed at 2.7 mW in order to simultaneously obtain an acceptable signal-to-noise ratio and avoid excessive heating of the graphene by the laser beam itself. For the recorded spectra the typical integration time was 60 seconds.

The data displayed in Figures 1 through 4 are recorded on one and the same device. Altogether three devices were measured. They showed the same generic features. Prior to the Raman experiments, the graphene devices were cleaned by current annealing[14] in order to remove molecular adsorbates and contaminants due to the fabrication procedure and exposure to the ambient. Apart from the source and drain contacts, the geometry includes voltage probes at either end of the constriction to monitor the voltage drop $V_C$ across the constriction. The current voltage characteristic $I$-$V_C$ at zero back-gate voltage is plotted in the right inset of Figure 1a. Note that the four-terminal configuration ensures that the calculated dissipated power $I \cdot V_C$ does not include the unsolicited power-dissipation due to the contact resistances at the source and drain. According to field effect measurements, charge neutrality is reached close to a back-gate voltage of -8 V which corresponds to an electron density of ~$5.7 \cdot 10^{11}$ cm$^{-2}$ and a Fermi energy of ~90 meV.

Figure 1a shows the Stokes and anti-Stokes Raman spectrum measured across the accessible spectral range for laser excitation at a wavelength $\lambda_0 = 633$ nm. It has been recorded at the center of the constriction for an electric power dissipation of 2.6 mW. Here, we will focus on the optical G-phonon Raman peak. It is attributed to an in-plane stretching mode of the carbon atoms[15] as shown in the cartoon between Figure 1b and c. Figure 1b illustrates the behavior of the anti-Stokes G-peak as a function of the dissipated electric power $I \cdot V_C$. The corresponding spectra for the Stokes G-peak are plotted in Figure 1c. From the intensity ratio of the Stokes ($I_S$) and anti-Stokes ($I_{As}$) G-peak it is possible to extract the occupation number of the G-phonons or alternatively the effective temperature $T_G$ describing their Bose-Einstein distribution probability using the following equation[16]:

$$\frac{I_{As}}{I_S} = \left( \frac{\omega_0 + \omega_G}{\omega_0 - \omega_G} \right)^4 \exp\left( -\frac{\hbar \omega_G}{k_B T_G} \right).$$ Here $\hbar \omega_0$, $\hbar \omega_G$, $\hbar$, and $k_B$ are the energy of the incident photons, the G-phonon energy, the Planck and Boltzmann constants. The temperature $T_G$ is obtained by fitting the Stokes and anti-Stokes lines to Lorentzians, taking into account the wavelength dependent sensitivity of the detection system for the Raman peaks[17], and calculating the intensity ratio according to the above equation for a photon energy $\hbar \omega_0 = 1.959$ eV. The temperature value has been included near each curve in Figure 1b and c. An anti-Stokes line can be observed starting from about 500 K. The absence of an anti-Stokes feature in the bottom trace of Figure 1b for zero electric power confirms that the incident laser light for recording the Raman data does not raise the local temperature of the G-phonons above 500 K. The temperature of the Si bulk substrate, located 300 nm below the flake, was calculated from the Si lines[9] at $\pm 520$ cm$^{-1}$. It was approximately 300 K independent of the incident laser power up to the maximum of 2.7 mW. At the highest electric power dissipation in the constriction, the temperature of the Si substrate underneath rises by about 10 K.

The spatial profile of $T_G$ along the constriction for 2.4 mW of electric power dissipation was obtained by recording Raman spectra as a function of position with a spatial resolution of 400 nm. The outcome of this experiment is depicted in Figure 2a. At the center of the constriction $T_G$ is raised above 1600 K, indicating that it is possible to generate a large non-equilibrium population of G-phonons with electrical biasing. Outside of the constriction region, $T_G$ is nearly position independent and equal to 750-800 K. By comparing the current density in this region to the measured data at the constriction we expect a temperature around 600 to 700 K. This indicates that the elevated temperature outside of the constriction is also mainly caused by Joule heating. A spatial map of the anti-Stokes G-peak intensity is plotted in Figure 2b. This color rendition was recorded on a 0.2 µm spatial grid with an integration time of 10 seconds for the same electric power dissipation of 2.4 mW. A "hot spot" is clearly visible at the constriction and confirms that power dissipation mainly takes place in this region. Figure 3a plots the G-phonon temperature as a function of the dissipated electric power at the center of the graphene constriction. Data points extracted from the Raman spectra displayed in Figure 1b and c are shown in the same color as these spectra. On another device experiments were carried out up to breakdown. When imposing a current density above ~1.4 mA/µm (this corresponds to ~4×10$^8$ A/cm$^2$ with a monolayer thickness of 0.35 nm) the constriction glows by emitting visible light (see Supporting Information). The maximum current density sustained by the graphene constriction prior to breakdown was 1.6 mA/µm (4.5×10$^8$ A/cm$^2$).

The intensities of the Stokes and anti-Stokes G-peak are affected by power dissipation, and also the position and the full width at half maximum (FWHM) of these Raman modes change. We first address the position of these Raman features. Figure 3b plots for instance the shift of the Stokes G-peak position $G_S$ as a function of the calculated G-phonon temperature $T_G$. $G_S$ shifts downward with increasing $T_G$ or equivalently with increasing $I \cdot V_C$. The temperature dependent shift in Figure 3b is qualitatively consistent with previous experimental results[18] and theoretical predictions[19]. According to theory[19], the thermal lattice contraction in graphene leads to a minor shift. Furthermore, this shift would be in the opposite direction. The dominant contribution to the line shift comes from anharmonic effects, which include three and four phonon scattering processes[19]. In covalent semiconductors[9], semimetals[11] as well as metals[10] also the anharmonic phonon contribution prevails. For the sake of completeness, we note that also the charge carrier density influences the position of the $G_s$-line[20, 21]. Adsorbates may be removed as the flake gets heated[22]. However, field effect measurements recorded after each experiment at a different value of $I \cdot V_C$ show that charge neutrality remains at the same back-gate voltage and no further dopants are removed. Most dopants were already removed during the annealing step prior to the experiments. Unintentional electrostatic gating when applying a large source-drain bias voltage may also modify the carrier density, however the effect was found to be negligible. Extended exposure to intense laser light may also produce changes in the G peak position[22]. To exclude that this effect occurred under the conditions of the present experiment, Raman spectra were compared at zero dissipated electric power before and after each measurement. No changes were observed. Quantitatively, the decrease of $G_S$ for a given $\Delta T_G$ is much smaller than for an equilibrium experiment in which an identical temperature change is accomplished by heating the entire substrate with electrical current. The data collected during such an experiment using another graphene flake are plotted in Figure 3c. The Stokes line red-shifts linearly at a rate of approximately 3.2 cm$^{-1}$ for a temperature change of 100 K. The discrepancy between the experiment with a heated substrate and the experiment with local power dissipation in the constriction reflects the non-equilibrium character of the G-phonon population near the constriction. It is instructive to determine the required equilibrium flake temperature in order to obtain the same $G_s$ line shift as for the non-equilibrium experiment. This temperature $T_G^{eq}$ has been plotted as the right ordinate in Figure 3b.

We now turn our attention to the behavior of the FWHM of the $G_s$-line with increasing electric power dissipation or equivalently $T_G$. The experimental data are summarized in Figure 4a. The FWHM slightly increases by 0.5 cm$^{-1}$ for a $T_G$ up to about 700 K. Then it decreases by as much as 2 cm$^{-1}$ and finally it starts to grow again for $T_G$ above ~1000 K. The anti-Stokes G-mode also shows a similar dependence. This non-monotonic temperature dependence of the G-linewidth is the key experimental observation. The decrease of the FWHM is distinct from the monotonous increase of the FWHM of optical phonons observed in semiconductors[9], other semi-metals[11], as well as metals[10]. As we will show below, this anomalous change of the G-linewidth is attributed to the electron-phonon coupling in graphene, which is able to compete with anharmonic phonon coupling up to high temperatures. The intrinsic FWHM of the G-mode describes the decay rate of the G-phonons[19]. This decay rate is composed of an electron-phonon and a phonon-phonon scattering contribution denoted as $\gamma_{e-ph}$ and $\gamma_{ph-ph}$. In principle also extrinsic broadening of the FWHM should be considered. For instance, the extrinsic broadening, $\gamma^*$ can be attributed to the inhomogeneity of the charge carrier density in the form of electron-hole puddles[23] or the spectral resolution limitation of the detection system. The final FWHM $\gamma$ can then be written as follows;

$\gamma = \gamma_{e-ph} + \gamma_{ph-ph} + \gamma^*$.

The electron-phonon coupling is described by Landau damping. G-phonons annihilate by creating electron-hole pairs. This process is possible even in the limit of absolute zero temperature, provided the chemical potential is positioned within an energy window, which is centered around the Dirac point and whose size is defined by the G-phonon energy (~0.2 eV). The decay rate $\gamma_{e-ph}$ of the G-phonons is given analytically from Fermi's golden rule in the vicinity of the **K** point of the Brillouin zone under the assumption of a linear energy-momentum dispersion of the charge carriers[24]

$$\gamma_{e-ph}(\mu, T_e) = \gamma^o_{e-ph}\left(f_{FD}\left(-\frac{\hbar\omega_G}{2}\right) - f_{FD}\left(\frac{\hbar\omega_G}{2}\right)\right), \qquad (1).$$

Here $f_{FD}(E) = \frac{1}{e^{(E-\mu)/k_B T_e} + 1}$ is the Fermi-Dirac distribution function. $\gamma^o_{e-ph}$, $\mu$, and $T_e$ denote the FWHM at zero temperature, the chemical potential, and the electronic temperature. When the chemical potential is located outside the energy interval $\left[-\frac{\hbar\omega_G}{2}, \frac{\hbar\omega_G}{2}\right]$, this decay process is prohibited (at least at absolute zero temperature) by the Pauli exclusion principle and the requirement of energy and momentum conservation. This Pauli blocking results in a long lifetime of phonons and, hence, a small FWHM. The charge carrier density enters $\gamma_{e-ph}$ via the chemical potential $\mu$. Recently, the influence of the carrier density on the FWHM of $G_s$-peak was verified with Raman studies by tuning the carrier density in a field-effect device[20, 21]. Here, however the density is fixed. As mentioned previously, the amount of chemical doping is not altered during the experiment and the influence of electrostatic gating as a result of the applied source-drain bias voltage remains limited even at the highest electric power dissipation. The observed behavior of the FWHM should therefore be attributed solely to a change of temperature.

In our Joule heated graphene constriction, the electronic temperature $T_e$ is unknown. $T_e$ will increase with increasing power dissipation, but the precise relation with $T_G$ remains inaccessible. Electrons and G-phonons are likely to adopt similar temperatures through rapid mutual scattering if electron-phonon coupling is strong[6, 25]. When $\mu$ is located close to the Dirac point, thermal broadening will reduce the number of occupied electronic states near $-\frac{\hbar\omega_G}{2}$ and also the number of empty states near $+\frac{\hbar\omega_G}{2}$. The probability for the generation of electron-hole pairs will drop accordingly with increasing $T_e$. This case is illustrated in the inset to Figure 4c. At higher temperatures, however, $\gamma_{ph-ph}$ will gain in strength and will eventually dominate[19]. In the simplest scenario, G-phonons will decay into two lower energy acoustic phonons[16]. More sophisticated higher order anharmonic processes were considered theoretically[19]. Phonon-phonon coupling produces a monotonic increase of the FWHM as the temperature grows. The competition between electron-phonon coupling and phonon-phonon coupling and the different temperature regimes where these mechanisms prevail can account for the observed non-monotonic behavior of the Raman $G_s$-linewidth.

Figure 4b plots the $\gamma_{\text{e-ph}}$ and $\gamma_{\text{ph-ph}}$ contributions separately, while Figure 4c shows the sum of both to illustrate that it is possible to obtain the qualitative behavior of the experimental data in Figure 4a. For simplicity, we consider an equilibrium case where the lattice or phonon temperatures and the electronic temperature are all equal (T). The red dashed curve represents $\gamma_{\text{e-ph}}$ according to equation (1). Here a $\gamma^o_{\text{e-ph}}$ of 11 cm$^{-1}$ was adopted as determined from density functional theory[19]. The chemical potential was calculated for charge neutrality at a back-gate voltage of -8 V. A temperature dependent shift of the chemical potential due to the thermal redistribution of charge carriers[26] was taken into account. This shift is not negligible in the temperature regime covered here (see Supporting Information). For the anharmonic contribution to the temperature dependent shape of the Raman linewidth, we restrict ourselves to an oversimplified model applicable when an optical phonon decays into two acoustic phonons[16]: $\gamma_{\text{ph-ph}} = \gamma^o_{\text{ph-ph}}\left(1 + \dfrac{2}{e^{\hbar\omega_G/2k_BT}-1}\right)$. The behavior described by this expression is plotted by the solid blue curve in Figure 4b. We have used 2 cm$^{-1}$ for $\gamma^o_{\text{ph-ph}}$ [19]. A theoretical treatment considering more additional phonon-phonon decay channels available in graphene was covered theoretically in Ref. 19. Despite the simplifying assumptions, the shape of the temperature dependent linewidth qualitatively agrees with the experimental data. Not only the drop of the linewidth, but also the small initial increase of the linewidth is captured by this model. The initial increase stems from phonon-phonon coupling. At intermediate temperatures, the thermal redistribution of charges suppresses the electron-phonon decay channel and the linewidth drops. At elevated temperatures, phonon-phonon coupling dominates again and imposes a linewidth increase as the temperature grows. We note that a drop of the linewidth upon raising the temperature is not always seen in graphene. This only occurs if the chemical potential remains close to the Dirac point and within the energy window of size $\hbar\omega_G$ centered around the Dirac point. For graphene doped at a level higher than $7\times10^{11}\text{cm}^{-2}$, corresponding to a chemical potential larger than $\frac{\hbar\omega_G}{2}$ the linewidth drop has vanished (see Supporting Information). This can easily be understood within the above model.

The linewidth drop in graphene is a dramatic manifestation of electron-phonon coupling. It occurs because of graphene's unique properties. It is gapless, the chemical potential can be tuned below half the optical phonon energy $\frac{\hbar\omega_G}{2}$ and local temperatures comparable to $\frac{\hbar\omega_G}{2}$ can easily be reached. In semiconductors, for instance, the bandgap usually exceeds the energy of optical phonons and electron-phonon decay remains inactive. In metals and other semimetals, the accessible temperature in experiment is usually two orders of magnitude below the Fermi temperature. Thus, it is not possible to access a regime where thermal redistribution of charge carriers would influence the decay process of optical phonons due to the electron-phonon coupling.

ACKNOWLEDGMENT. We thank Dong-Su Lee for helpful comments on the manuscript.

**Supporting Information Available:**

2D Raman peak of the measured graphene device, current-voltage characteristics and CCD image for a glowing graphene constriction, temperature dependence of the chemical potential, and temperature dependence of the G-linewidth for highly doped graphene. This material is available free of charge via the Internet at http://pubs.acs.org.

**Figures and Figure Captions:**

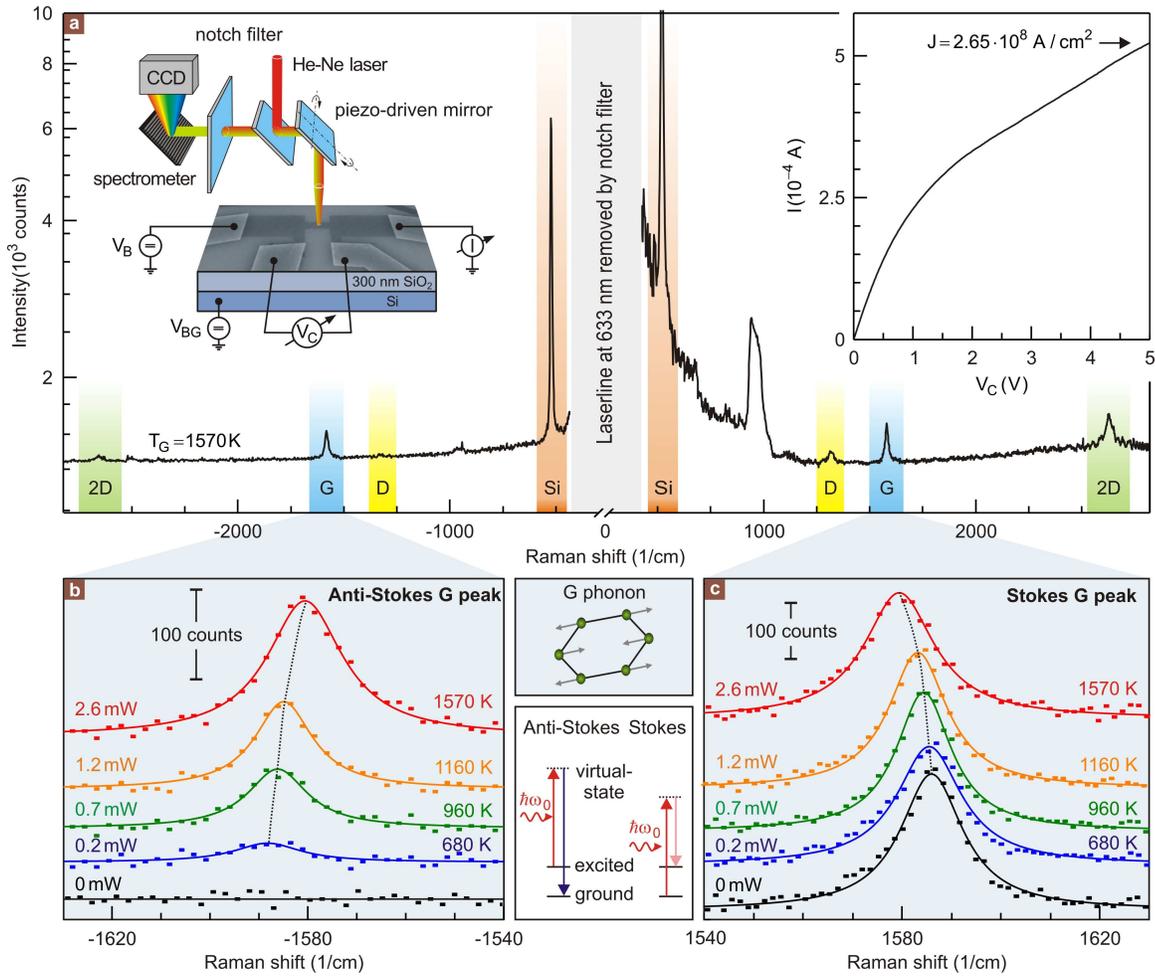

**Figure 1: Raman spectra of an electrically biased graphene constriction. a.** Stokes and anti-Stokes Raman spectra taken when 2.6 mW of electrical power is dissipated in the constriction. The prominent Raman peaks of graphene (G, D, and 2D modes) are marked as well as the Si peak around 520 cm$^{-1}$. The left inset depicts the measurement arrangement. It consists of a scanning confocal Raman spectrometer and a typical graphene constriction with a width of 0.6 μm and a length of 1.5 μm. The right inset shows the current-voltage characteristic measured at zero back-gate voltage across the constriction. **b.** Anti-Stokes and **c.** Stokes peaks of the G-mode acquired for zero up to 2.6 mW of dissipated electric power at the constriction. The spectra are fitted with Lorentzians (solid lines). They were shifted vertically for clarity. The effective temperature of the G-phonon was extracted from the intensity ratio of the Stokes

and anti-Stokes lines. This temperature has been included near each curve. The upper central inset displays the in-plane stretching eigenmode (G-phonon) in graphene. The bottom inset contains schematic diagrams for the Stokes and anti-Stokes Raman scattering processes.

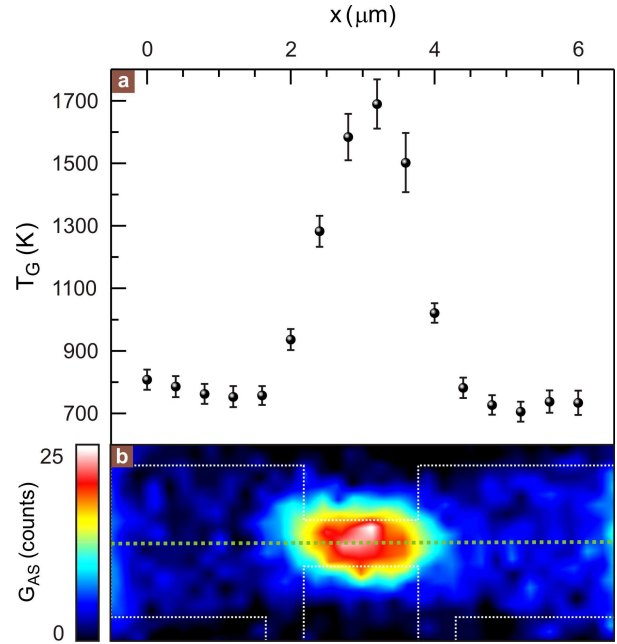

**Figure 2: Spatial profile of the effective G-phonon temperature $T_G$ across the device and the anti-Stokes spatial Raman map.** a. Cross-sectional profile of the effective temperature of the G-phonons $T_G$ at a dissipated-power of 2.4 mW. The effective temperature at the center is significantly higher than in the wide graphene regions close to the electrodes. The maximum value is larger than 1600 K. b. Color map of the anti-Stokes G-mode intensity acquired at a dissipated-power of 2.4 mW. Red (blue) corresponds to a high (low) intensity value and hence high (low) temperature. The white dotted line demarcates the etched graphene flake.

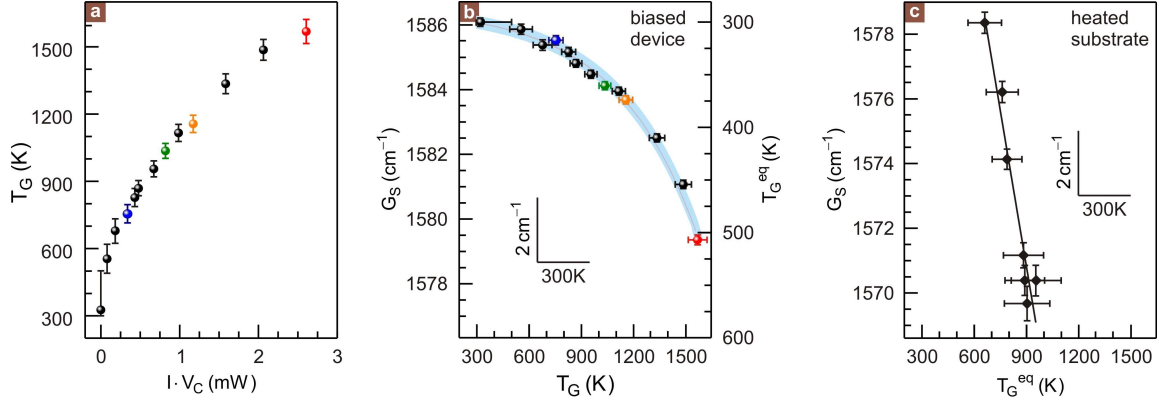

**Figure 3: Power dependence of the effective G-phonon temperature and Raman shift of the Stokes G-mode as a function of $T_G$ or $T_G^{eq}$.** a. Effective temperature of the G-phonons at the center of the graphene constriction as a function of the dissipated electrical power $I \cdot V_C$. b. Shift of the $G_S$-line as a function of $T_G$ (left and bottom axes) for local heating at the constriction. $T_G$ is determined from panel a. $T_G^{eq}$ (right ordinate, determined from panel c) gives the equilibrium temperature required to obtain the same $G_S$ shift. The color-coded circular data points in a and b are extracted from the Raman spectra displayed in Figure 1b and c. c. Shift of the Stokes G-peak as a function of the equilibrium temperature $T_G^{eq}$. In this experiment, the entire sample is heated by driving an electrical current through the conducting Si substrate and the temperature is determined from the Stokes to anti-Stokes ratio of the G peak.

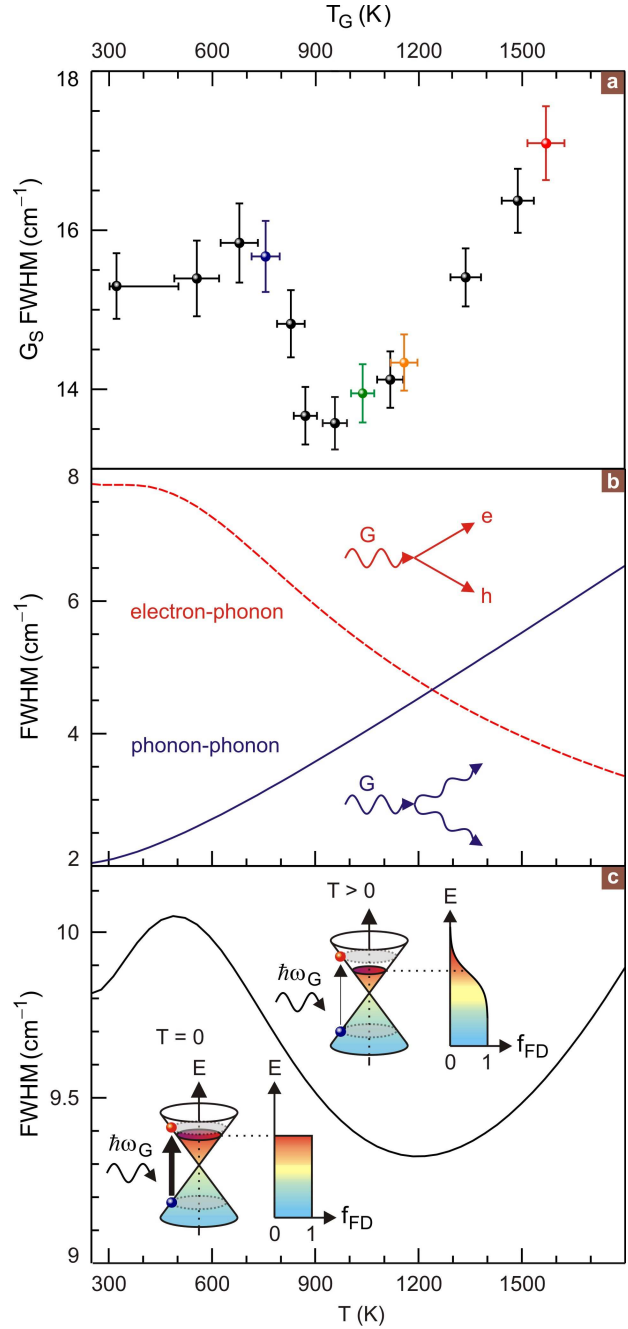

**Figure 4: Linewidth of the Stokes G-mode as a function of the effective G-phonon temperature. a.** $T_G$ dependence of the FWHM of the Stokes G-mode. The FWHM depends on $T_G$ in a non-monotonic fashion. The color-coded data points are obtained from the Raman spectra plotted in the same colors in Figure 1b and c. **b.** The electron-phonon (dashed red line) and phonon-phonon (solid blue line) decay contributions to the intrinsic FWHM of the G-mode as a function of temperature. The electron-phonon

contribution was calculated using equation (1). For the phonon-phonon contribution only the simplest anharmonic process was considered (see text). No back-gate voltage is applied and charge neutrality is reached at -8V back-gate voltage. **c**. The sum of the above two contributions qualitatively predicts the non-monotonic $T$ dependence of the FWHM seen in experiment. Note that the extrinsic broadening of the linewidth due to for instance density inhomogeneity and spectral resolution was not taken into consideration here. The right inset shows the thermal broadening of the Fermi-Dirac distribution at higher temperatures while the left inset shows the corresponding distribution at zero temperature. A temperature increase suppresses the decay of G-phonons into electron-hole pairs.

**References**

1. Yao, Z.; Kane, C. L.; Dekker, C.  Phys. Rev. Lett. 2000, 84, 2941.

2. Park, J. Y., et al.  Nano Lett. 2004, 4, 517.

3. Bushmaker, A. W.; Deshpande, V. V.; Bockrath, M. W.; Cronin, S. B.  Nano Lett. 2007, 7, 3618.

4. Oron-Carl, M.; Krupke, R.  Phys. Rev. Lett. 2008, 100, 127401.

5. Steiner, M., et al.  Nature Nanotech. 2009, published online 1 March 2009.

6. Lazzeri, M.; Mauri, F.  Phys. Rev. B 2006, 73, 165419.

7. Pop, E., et al.  Phys. Rev. Lett. 2005, 95, 155505.

8. Meric, I., et al.  Nature Nanotech. 2008, 3, 654.

9. Lang, G., et al.  Phys. Rev. B 1999, 59, 6182.

10. Schulz, H.; Hüfner, S.  Solid State Commun. 1976, 20, 827.

11. Höhne, J.; Wenning, U.; Schulz, H.; Hüfner, S.  Z. Physik B 1977, 27, 297.

12. Novoselov, K. S., et al.  Science 2004, 306, 666.

13. Ferrari, A. C., et al.  Phys. Rev. Lett. 2006, 97, 187401.

14. Moser, J.; Barreiro, A.; Bachtold, A.  Appl. Phys. Lett. 2007, 91, 163513.

15. Reich, S.; Thomsen, C.  Phil. Trans. R. Soc. London. A 2004, 362, 2271.

16. Kuzmany, H. Solid-State Spectroscopy; Springer-Verlag: Berlin, 1998.

17. Correction factors were estimated for each wavelength by comparing the spectrum from a calibrated tungsten light source with that of the theoretical black body curve.
17

# Hot phonons in an electrically biased graphene constriction

Dong-Hun Chae, Benjamin Krauss, Klaus von Klitzing, and Jurgen H. Smet

Max-Planck-Institute for Solid State Research, Heisenbergstrasse 1, 70569 Stuttgart, Germany

**Supporting Information**

2D Peak of the measured graphene device

Figure S1 depicts the 2D Raman spectrum of the measured graphene device. The line-shape and position clearly reveals that the device is fabricated out of a monolayer.

Maximum current density of a graphene constriction

Figure S2 depicts the $I-V$ characteristics of another device with the same constriction geometry as the device described in the main text (a constriction width of 0.6 μm and a length of 1.5 μm). This device has only two terminals. The current through the device eventually saturates at high bias voltage and the graphene constriction burns out as signaled by the sharp current drop. The current density at breakdown is larger than 1.6 mA/μm ($4.5 \times 10^8$ A/cm$^2$). This is comparable to the high current densities reported for carbon nanotubes[1]. During the acquisition of the current voltage characteristic, pictures of the device were captured with a CCD camera and white source illumination. The left inset shows an initial reference snapshot taken with low power applied to the device whereas the upper right inset is with a high applied voltage. The lower right inset schematically shows the device with the electric contacts (yellow) and the etched graphene constriction (blue). The right inset shows that the constriction glows when applying a bias voltage of about 10 V. The constriction emits visible radiation above a current density of ~1.4 mA/μm (~$4 \times 10^8$ A/cm$^2$).

Temperature dependence of the chemical potential

At high temperatures, the thermal redistribution of charges causes a non-negligible shift of the chemical potential. This change of the chemical potential needs to be taken into account when determining the temperature dependence of the electron-phonon decay using Eq. 1. In order to calculate the chemical potential we assume that the net charge carrier density remains fixed at $q_o$ independent of temperature, $n - p = q_o$, where $n$ and $p$ follow from

$$n = \int_0^\infty \rho(\varepsilon) f_{FD}(\varepsilon, \mu, T_e) d\varepsilon = -2 \frac{k_B^2 T_e^2}{\pi \hbar^2 v_F^2} Li_2\left(-e^{\mu/k_B T_e}\right)$$ and

$$p = \int_{-\infty}^0 \rho(\varepsilon)(1 - f_{FD}(\varepsilon, \mu, T_e)) d\varepsilon = 2 \frac{k_B^2 T_e^2}{\pi \hbar^2 v_F^2} Li_2\left(-e^{-\mu/k_B T_e}\right).$$

Here, $\rho$ is the density of states for graphene and $f_{FD}$ is the Fermi-Dirac distribution function. $Li_2(x)$ stands for the dilogarithm polynomial function $Li_2(x) = \sum_{j=1}^\infty (x^j / j^2)$. Figure S3 depicts the temperature dependence of the chemical potential for $q_o = 5.7 \times 10^{11}$ cm$^{-2}$. This corresponds to the net charge carrier density of the graphene flake used to obtain the data of the main text (charge neutrality is reached at -8 V back-gate voltage and measurements were carried out at 0 back-gate voltage).

Temperature dependence of the G-linewidth for highly doped graphene

Figure S4a shows the G-linewidth as a function of $T_G$ for a graphene constriction with a larger amount of intrinsic doping (blue diamonds). Charge neutrality in this device is reached at a back-gate voltage of -15 V. Zero back-gate voltage corresponds to a carrier density of approximately $1 \times 10^{12}$ cm$^{-2}$ and the chemical potential at room temperature is larger than $\frac{\hbar \omega_G}{2} \cong 100$ meV. The linewidth no longer shows the clear drop with increasing $T_G$. The decay of G-phonons through the creation of electron-hole pairs is suppressed as a result of the Pauli exclusion principle and the need for energy and momentum conservation. For comparison show the red circles in Fig. S4a a situation where charge neutrality was reached at -8 V. The linewidth now exhibits a pronounced drop in the intermediate temperature regime.

This experiment confirms that electron-phonon coupling only dominates for small enough carrier densities such that the absolute value of the chemical potential is located below $\frac{\hbar\omega_G}{2}$.

In Fig. S4b and c we consider theoretically the temperature dependent behavior of the G-linewidth as a function of the doping level within the framework outlined in the main text. Fig. S4b shows the electron-phonon scattering contribution for different values of the carrier density (colored lines) and the phonon-phonon decay contribution (black line) (see also Fig. 4 of the main text) to the G-linewidth. The carrier density is specified in terms of the Fermi energy, i.e. the chemical potential at absolute zero temperature. Figure S4(c) plots the sum of the two contributions. If the Fermi energy is close to the Dirac point, the temperature dependence of the linewidth is dominated by a gradual suppression of the electron-phonon scattering with increasing temperature. In the opposite regime for large carrier densities, the linewidth monotonously increases.

Figure S1

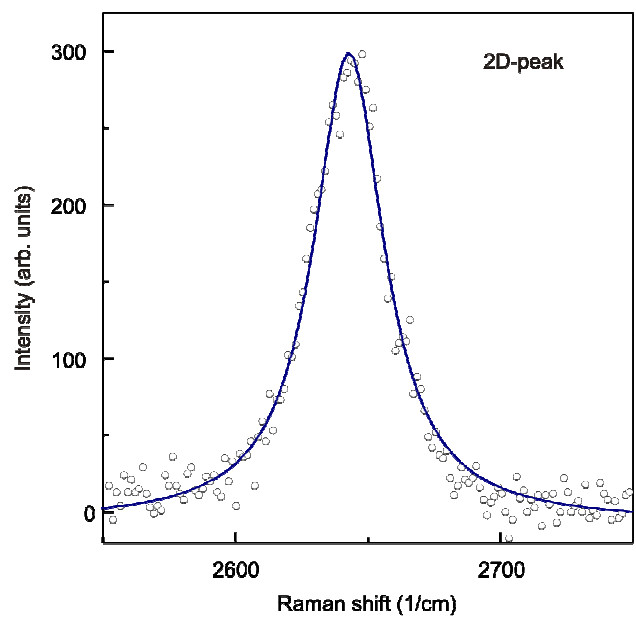

Figure S2

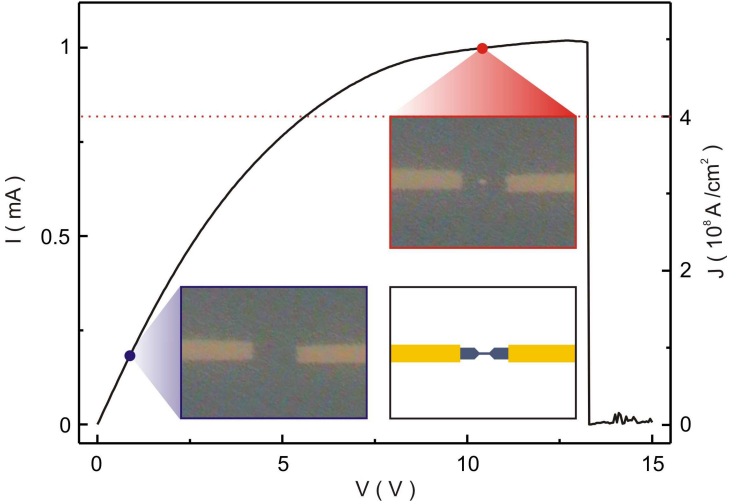

Figure S3

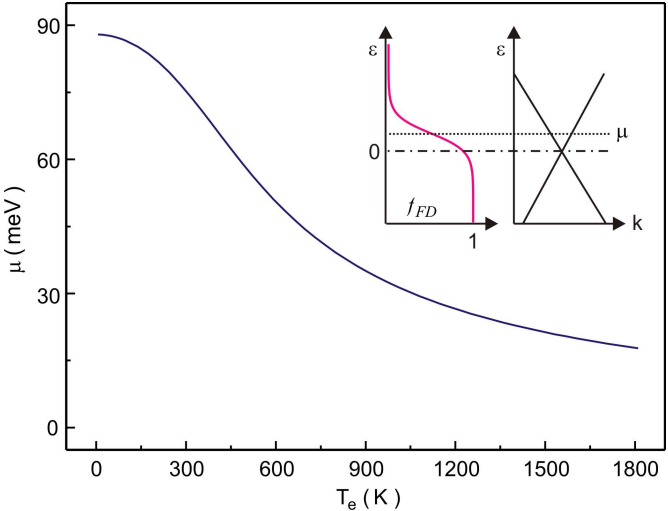

Figure S4

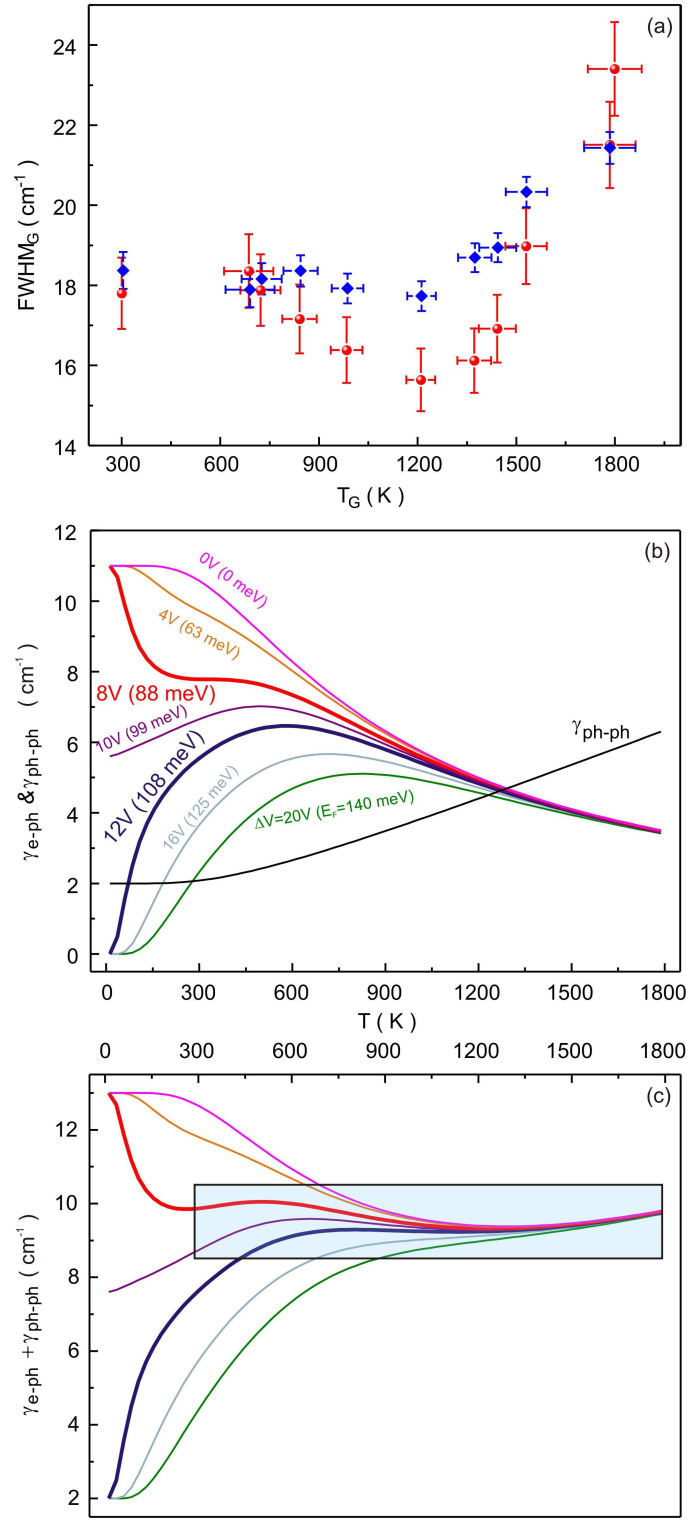